# From Flat to Spatial: Comparison of 4 methods constructing 3D, $2^{1/2D}$ Models from 2D Plans with neural networks.

Jacob Sam[1], Karan Patel[2], Mike Saad[3]

**Abstract**

In the field of architecture, the conversion of single images into 2½D and 3D meshes is a promising technology that enhances design visualization and efficiency. This paper evaluates four innovative methods: "One-2-3-45," "CRM: Single Image to 3D Textured Mesh with Convolutional Reconstruction Model," "Instant Mesh," and "Image-to-Mesh." These methods are at the forefront of this technology, focusing on their applicability in architectural design and visualization. They streamline the creation of 3D architectural models, enabling rapid prototyping and detailed visualization from minimal initial inputs, such as photographs or simple sketches.One-2-3-45 leverages a diffusion-based approach to generate multi-view reconstructions, ensuring high geometric fidelity and texture quality. CRM utilizes a convolutional network to integrate geometric priors into its architecture, producing detailed and textured meshes quickly and efficiently. Instant Mesh combines the strengths of multi-view diffusion and sparse-view models to offer speed and scalability, suitable for diverse architectural projects. Image-to-Mesh leverages a generative adversarial network (GAN) to produce 3D meshes from single images, focusing on maintaining high texture fidelity and geometric accuracy by incorporating image and depth map data into its training process. It uses a hybrid approach that combines voxel-based representations with surface reconstruction techniques to ensure detailed and realistic 3D models.This comparative study highlights each method's contribution to reducing design cycle times, improving accuracy, and enabling flexible adaptations to various architectural styles and requirements. By providing architects with powerful tools for rapid visualization and iteration, these advancements in 3D mesh generation are set to revolutionize architectural practices.

***Keywords:*** *Applied Research, scale XL, Image to 3d, Mesh Reconstruction, Generative Design*

**Introduction**

The evolution of digital tools within the architectural domain has significantly reshaped how architects engage with the design and visualization of structures. The capacity to translate two-dimensional imagery into 21/2d,3d models is not merely a technical advancement but a paradigm shift that augments the creative process, enhances client communication and refines the realization of architectural concepts. As the demand for faster and more accurate visual representations grows, the development of methods that facilitate the quick conversion of single images into detailed 3D meshes is becoming increasingly vital. These technologies are crucial in environments where rapid iteration and visualization are required, from preliminary design phases to the final presentation stages. (Yi et al. 2023)(2024)(Sun et al. 2018)

Architectural visualization has traditionally relied on physical models and CAD drawings, which can be time-consuming and resource-intensive. However, recent advancements in single-image 3D mesh generation technologies offer a more efficient way to create detailed 3D models (Kanazawa et al., 2018). These technologies enable architects to quickly iterate and explore various design options without the usual overhead associated with 3D modeling (Song et al., 2017). By using single images, these methods can automatically estimate 3D shapes, cameras, and textures of objects, significantly reducing the manual effort required (Pan et al., 2019). The emergence of deep learning methods has further enhanced the capabilities of 3D mesh generation. Techniques like "Deep Mesh Reconstruction From Single RGB Images via Topology Modification Networks" have shown superior performance in generating 3D shapes with complex topologies (Pan et al., 2019). Additionally, approaches such as "Pixel2Mesh: Generating 3D Mesh Models from Single RGB Images" have introduced innovative ways, like using geometry images, to represent 3D shapes efficiently (Wang et al., 2018). Moreover, the field has seen advancements in completing and labeling 3D scenes, as demonstrated in works like "Semantic Scene Completion from a Single Depth Image" (Song et al., 2017). These developments have streamlined the process of creating accurate and detailed 3D visualizations, especially in the early design stages where quick modifications are essential.

This paper explores four advanced approaches for single-image to 3D mesh conversion: "Image-to-Mesh," "One-2-3-45," CRM, and "Instant Mesh." "Image-to-Mesh" leverages advanced deep learning techniques to convert single images into detailed 3D meshes directly, balancing computational efficiency and output quality. "One-2-3-45" employs a sophisticated diffusion model to synthesize multiple views from a single input image, facilitating high-fidelity 3D reconstructions that maintain the original architectural intent(Liu et al. 2023). The CRM method enhances textural and structural detail by integrating geometric priors into its convolutional network, crucial for accurate architectural representation. "Instant Mesh" combines Multiview diffusion with sparse-view large reconstruction models to produce high-quality meshes rapidly, catering to diverse scales and complexities of architectural projects. These methods collectively represent significant advancements in the field, each contributing uniquely to the evolving landscape of 3D modeling from a single image. (2024)(Yi et al. 2023)

In examining these methodologies, the paper aims to highlight how each contributes to reducing turnaround times, improving geometric accuracy, and providing flexible tools that adapt to

*"From Flat to Spatial: Comparison of 4 methods
constructing 3D, 21/2D Models from 2D Plans
with neural networks"*

various architectural styles and project requirements. This exploration is intended to guide architects and designers in selecting appropriate tools that align with their specific workflow needs, thereby optimizing the design process and enhancing the overall quality of architectural outputs.

**Methodology**

The methodology began with the creation of a dataset using the Pix2Pix algorithm, a conditional generative adversarial network (cGAN), to process and enhance architectural floor plans. The dataset was compiled from existing architectural floor plans and enriched with an additional dataset of crystal structures to introduce visual complexity. The Pix2Pix algorithm learned from these datasets to generate enhanced 2D images, which served as the basis for further 2.5D and 3D visualization using four different models.

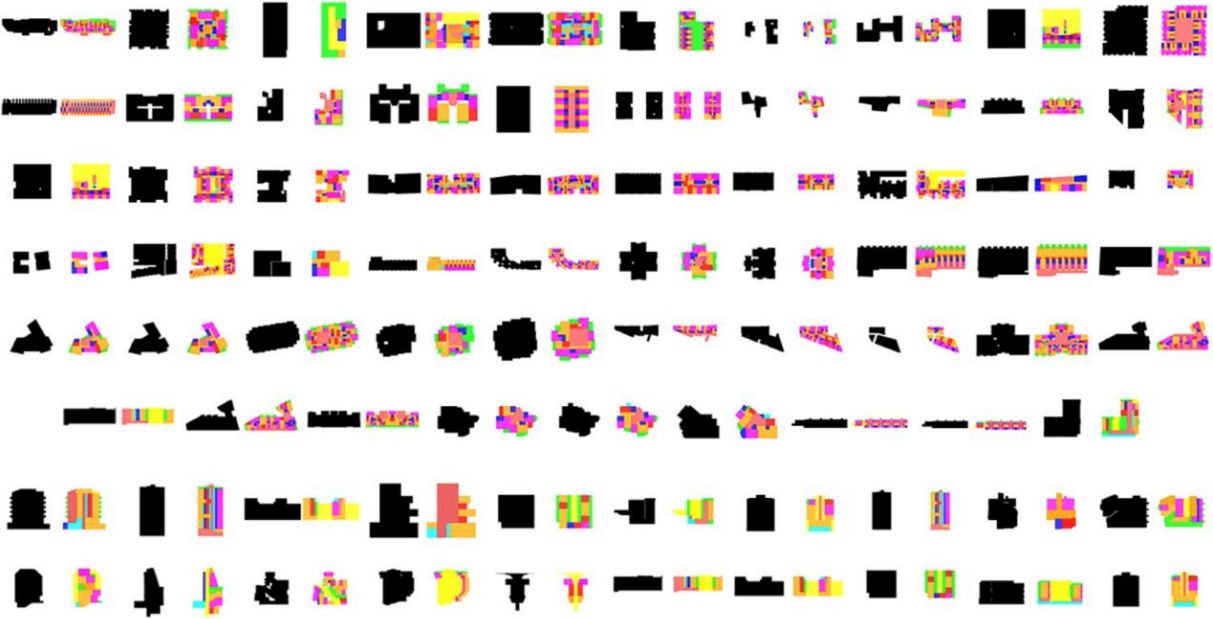

*Figure 1 dataset of floor plans*

*"From Flat to Spatial: Comparison of 4 methods constructing 3D, 21/2D Models from 2D Plans with neural networks"*

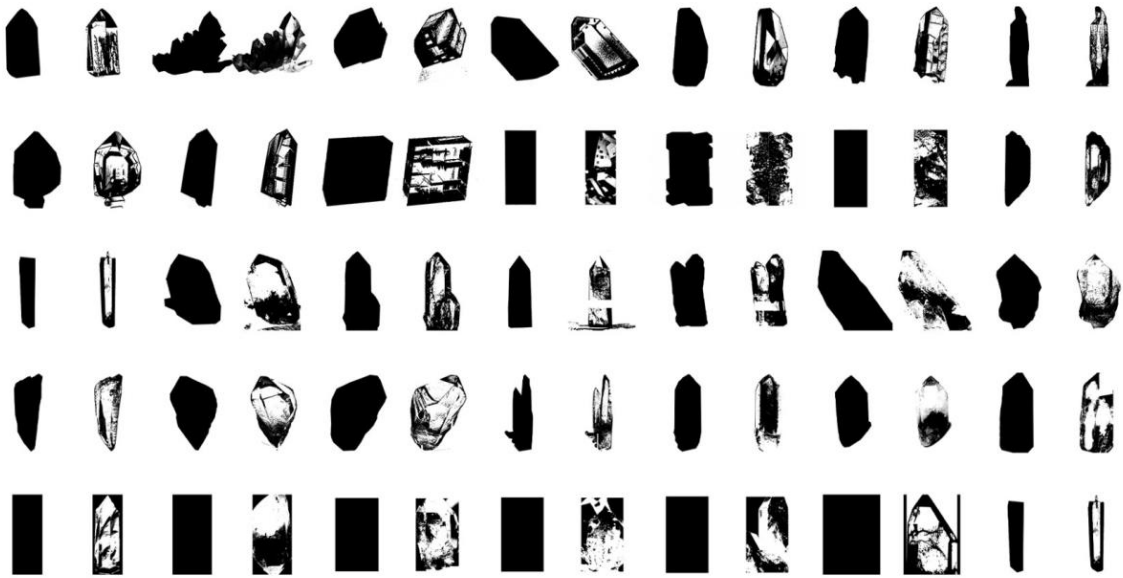

*Figure 2 dataset of crystal*

The techniques used in the work are described as follows:

- **Image-to-Mesh**: This approach leverages a generative adversarial network (GAN) to produce 3D meshes from single images. It focuses on maintaining high texture fidelity and geometric accuracy by incorporating both image and depth map data into its training process. It utilizes a hybrid approach that combines voxel-based representations with surface reconstruction techniques to ensure detailed and realistic 3D models.

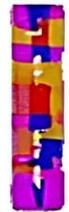 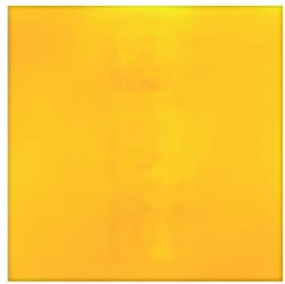 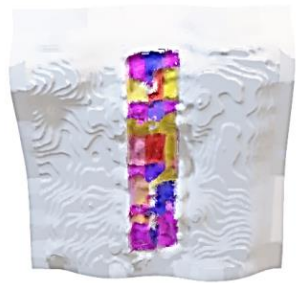

Input Image    Depth Map    Output Model

*"From Flat to Spatial: Comparison of 4 methods constructing 3D, 21/2D Models from 2D Plans with neural networks"*

- **One-2-3-45**: This method utilizes a view-conditioned 2D diffusion model named Zero123 for generating multi-view images from a single input. It employs a generalizable neural surface reconstruction method based on signed distance functions (SDF) to create 360-degree 3D meshes and incorporates several critical training strategies to enhance mesh geometry and consistency.

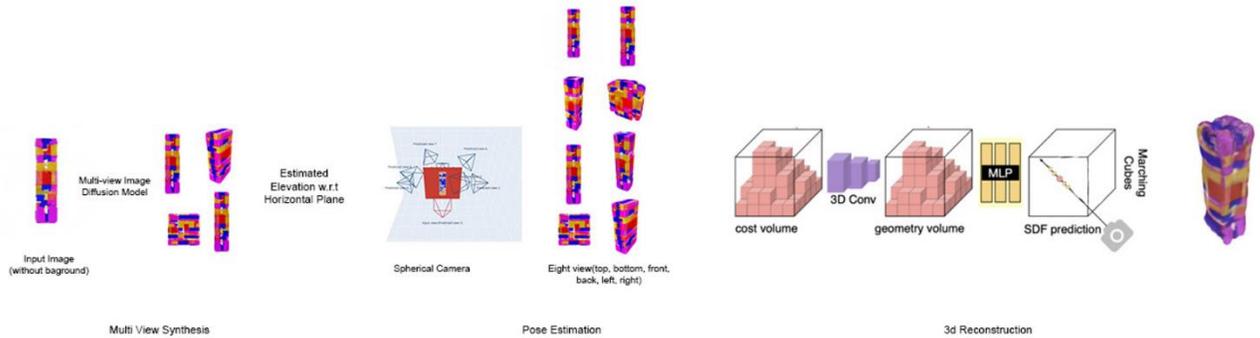

- **CRM (Convolutional Reconstruction Model)**: This model generates six orthographic view images using a convolutional U-Net, capitalizing on its strong pixel-level alignment capabilities. It employs a geometric representation known as Flexicubes, optimized directly on textured meshes to improve the fidelity of the output.

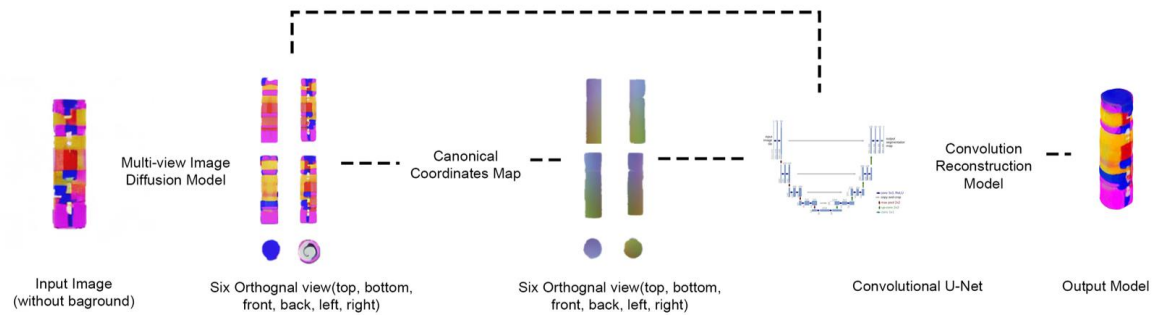

- **Instant Mesh**: This method combines an off-the-shelf Multiview diffusion model with a sparse-view reconstruction model based on the LRM architecture. It integrates a differentiable iso-surface extraction module for enhancing training efficiency and geometric supervision.

*"From Flat to Spatial: Comparison of 4 methods constructing 3D, 21/2D Models from 2D Plans with neural networks"*

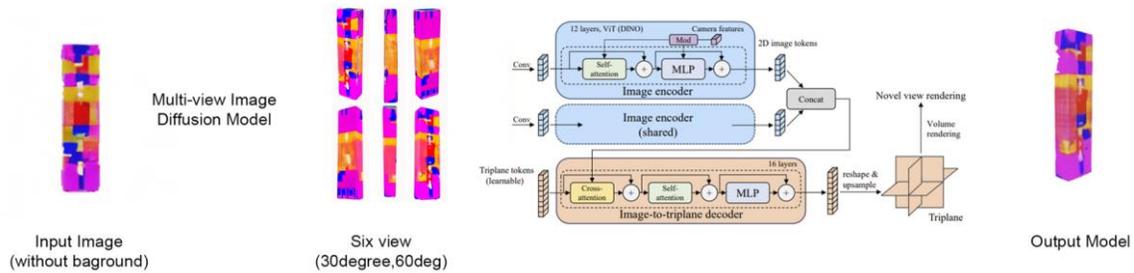

## Evaluation

The evaluation of each method was conducted rigorously to assess their effectiveness in architectural design contexts. The evaluation of the four neural network-based methods—"One-2-3-45,"Image to Mesh", "CRM (Convolutional Reconstruction Model)," and "Instant Mesh"—was meticulously designed to assess their utility in architectural visualization through both quantitative metrics and qualitative evaluations.

## Quantitative Metrics:

1. **Model Accuracy:** This metric assesses how closely the 3D models produced by each method resemble the original 2D plans and any existing structures. Accuracy can be quantified using error rate measurements, such as mean squared error (MSE) for geometrical features and structural alignment, providing a clear numerical benchmark for each method's performance.

2. **Efficiency:** Efficiency is evaluated by measuring the time and computational resources required to convert 2D images into 3D models. This includes processing time, measured in seconds or minutes, and resource utilization, such as CPU and GPU loads. Efficiency metrics help determine the practicality of each method in a professional workflow, particularly under tight deadlines or large-scale projects.

3. **Detail Resolution:** The resolution of details in the final models is quantified by examining the sharpness of textures and the precision of minor architectural elements. Methods like edge detection and pixel clarity analysis provide numerical values that reflect the ability of each method to preserve fine details from the original plans into the 3D models.

*"From Flat to Spatial: Comparison of 4 methods*
*constructing 3D, 21/2D Models from 2D Plans*
*with neural networks"*

**Qualitative Evaluations:**

1. **Aesthetic Quality:** This evaluation involves subjective assessments of the visual appeal and realism of the 3D models. Factors considered include texture quality, lighting accuracy, and overall visual integration into potential real-world settings. A panel of architectural experts might rate these aspects, providing qualitative feedback that complements the quantitative metrics.

2. **Architectural Performance:** Architectural performance looks beyond aesthetics, focusing on the functionality and spatial dynamics of the models. This includes evaluating how well the spaces within the models serve their intended purpose, their compliance with architectural standards, and their adaptability to real-world modifications. This can be assessed through scenario-based testing where models are critiqued based on their practical application in real-life architectural projects.

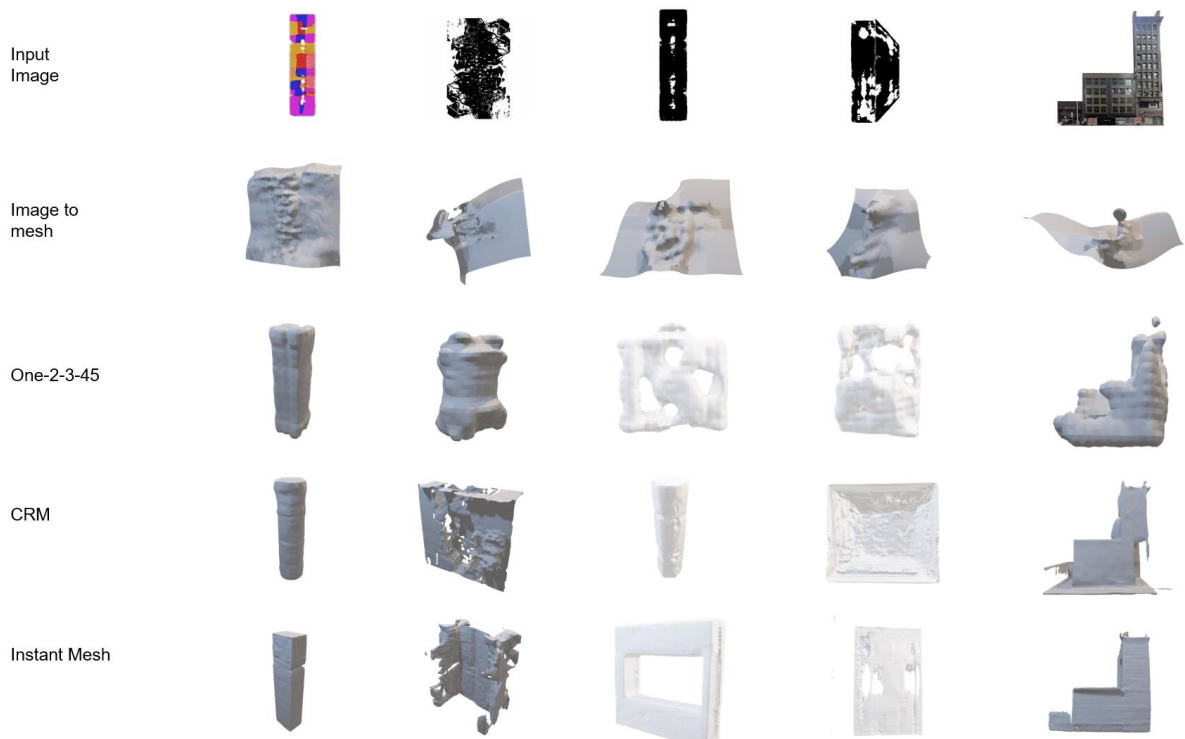

*Figure 3 Evaluation of the models*

*"From Flat to Spatial: Comparison of 4 methods constructing 3D, 21/2D Models from 2D Plans with neural networks"*

**3.Innovation and Creativity:** This aspect evaluates the novelty and originality of the solutions each method provides. It considers the extent to which these methods introduce new capabilities or significantly improve upon existing ones. Innovation can be assessed by comparing the methods against current industry standards to see if they push the boundaries of what's currently possible in architectural modeling.

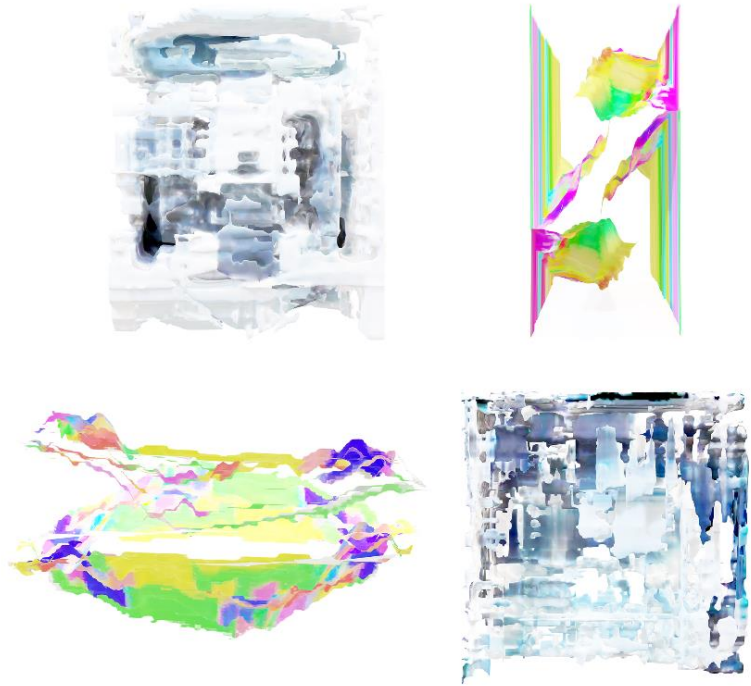

*Figure 4 Creative outputs from using these different models*

**Summary of Findings:**

This research explored three advanced neural network-based methods for converting 2D architectural plans into 3D and 2.5D models. Our findings highlight significant advancements in the field of architectural visualization:

- **One-2-3-45** demonstrated high geometric fidelity and texture quality, making it particularly suitable for projects where visual accuracy is paramount.

- **Image-to-mesh** is beneficial as it enables the creation of topological surfaces directly from images. This capability streamlines the process of converting visual data into usable 3D models.

*"From Flat to Spatial: Comparison of 4 methods*
*constructing 3D, 21/2D Models from 2D Plans*
*with neural networks"*

- **CRM (Convolutional Reconstruction Model)** proved to be highly efficient in generating textured meshes, with its use of Flexicubes enhancing the fidelity of output, thus favoring projects requiring rapid prototyping.
- **Instant Mesh** combined scalability with speed, ideal for handling diverse architectural projects due to its integration of Multiview and sparse-view diffusion models.

Each method has shown a potential to significantly reduce design cycle times, improve model accuracy, and provide flexible solutions adaptable to various architectural styles and requirements.

**Future Work:**

While this study has made substantial contributions to the field of architectural modeling, several avenues for future research have been identified:

- **Improvement of Data Sets:** Further enhancement and diversification of training datasets could improve model robustness and output quality across more varied architectural styles.
- **Real-Time Processing:** Developing methods for real-time model generation could revolutionize client presentations and iterative design processes.
- **Integration with Virtual and Augmented Reality:** Exploring the integration of these models into VR and AR platforms could provide more immersive and interactive design experiences.
- **Environmental and Structural Simulation:** Future work could also incorporate environmental impact analyses and structural integrity simulations directly tied to the generated models, offering a more comprehensive design tool.

**Final Thoughts:**

The journey from 2D floor plans to 3D architectural models represents more than a technological leap; it signifies a paradigm shift in how architects and designers conceptualize and interact with their creations. The methods discussed in this paper not only enhance the visual and functional aspects of architectural design but also promise to redefine the boundaries of architectural creativity and innovation. As we continue to refine these technologies, the future of architectural design appears not only more efficient but also more dynamic and responsive to the needs of both creators and users. Ultimately, the continued development of these tools will lead to smarter, more adaptable building designs that better serve their communities.

*"From Flat to Spatial: Comparison of 4 methods*
*constructing 3D, 21/2D Models from 2D Plans*
*with neural networks"*

*"From Flat to Spatial: Comparison of 4 methods constructing 3D, 21/2D Models from 2D Plans with neural networks"*

*"From Flat to Spatial: Comparison of 4 methods
constructing 3D, 21/2D Models from 2D Plans
with neural networks"*

"To improve the clarity and grammatical accuracy of the textural content, we utilized the capabilities of ChatGPT [https://chat.openai.com/] for copy-editing purposes only.

*"From Flat to Spatial: Comparison of 4 methods constructing 3D, 21/2D Models from 2D Plans with neural networks"*